# Energy Efficiency of Server-Centric PON Data Center Architecture for Fog Computing

Sanaa Hamid Mohamed, Taisir E. H. El-Gorashi, and Jaafar M. H. Elmirghani
*School of Electronic & Electrical Engineering, University of Leeds, LS2 9JT, United Kingdom*

**ABSTRACT**

In this paper, we utilize Mixed Integer Linear Programming (MILP) models to compare the energy efficiency and performance of a server-centric Passive Optical Networks (PON)-based data centers design with different data centers networking topologies for the use in fog computing. For representative MapReduce workloads, completion time results indicate that the server-centric PON-based design achieves 67% reduction in the energy consumption compared to DCell with equivalent performance.

**Keywords**: Data Center Networking (DCN), Passive Optical Networks (PON), Optical Line Terminal (OLT), MapReduce, energy efficiency, completion time.

## 1. INTRODUCTION

Driven by the increasing storage, processing, and networking demands of Internet of Things (IoT) devices and the strict latency requirements of their applications, several computing and storage solutions at edge and access networks were proposed to assist the cloud computing infrastructures at the core network [1]-[4]. These solutions utilize a wide range of devices ranging from switches, gateways, and access points [3], to wirelessly connected small data centers (i.e. cloudlets) [4]. In addition to benefiting the applications through reduced latency, these solutions also aid in reducing the congestion and improving the energy efficiency of core networks with cloud computing services and applications [5]-[11] under the increasing traffic volumes.

Several electronic, hybrid, and optical Data Center Networks (DCN) were proposed to connect the servers in cloud infrastructures offering different performance, power consumption, scalability, and cost [12]. This comes as a result of utilizing different technologies such as modern and commodity electronic switches, Network Interface Cards (NIC) attached to servers, and various passive and active optical components and devices for the interconnections [13]. As with cloud networking infrastructures, the increasing demands of data processing are also challenging the DCN because most of big data processing applications require extensive all-to-all server communications due to their distributed nature [14]. The impact of state-of-the-art DCN topologies on the performance and/or the energy efficiency of big data applications have been considered in [15]-[18]. To overcome the increasing power consumption and congestions in current data centers, and to meet the heterogeneous performance requirements of big data applications, an increasing number of hybrid and all-optical DCNs are proposed. Motivated by the energy efficiency, low cost, and high performance of Passive Optical Networks (PON) in access networks, different PON components were utilized for DCNs either to assist in hybrid electronic/optical deployments by connecting Top of Rack (ToR) electronic switches [19]-[21], or for all optical topologies [22]-[27]. In [22], five novel all-optical DCN topologies based on passive optical and PON technologies were designed. The third design, discussed in [23], provides high performance switching-centric interconnections by utilizing passive optical components (i.e. 1:N and N:N Arrayed Waveguide Grating Routers (AWGR), splitters, couplers, and passive polymer backplanes [28]), in addition to tunable optical transceivers or tunable Optical Network Unit (ONU) attached to servers. An optimization of the wavelengths assignment for different inter-rack links while considering the routing constraints of AWGRs is presented in [23]. Furthermore, the energy efficiency of the proposed design was compared to Fat-Tree and BCube, and energy savings of 45% and 80% were achieved, respectively. The fifth design is a cost-efficient server-centric architecture that eliminates the need for expensive tunable optical transceivers by utilizing Network Interface Cards (NIC) with non-tunable optical transceivers in the servers for inter-rack links [24]. Optimization results for the energy efficiency when considering the selection of relaying servers and the provisioning of servers computing resources showed average energy savings of up to 59% compared to random workloads placement. The fourth design which combines the benefits of the switching-centric and the server-center designs is optimized in [25]. Further resources provisioning optimizations were performed in [26] for different PON-based DCNs. In [27], the challenges associated with controlling the routing is addressed by facilitating the capabilities of Software Defined Networking (SDN) to control the optical layer [29]. The energy efficiency and cost effectiveness of PON equipment suggest the use of PON-based DCN not only in massive cloud data centers [23], but also in existing PONs at the edge of the network for fog computing as considered in [30]-[32]. Small scale PON-based DCNs can be attached in the access network for extended processing capabilities.

In this paper, we compare the energy consumption and completion time for MapReduce sort workloads as in our previous work in [18] while additionally considering the server-centric PON-based DCN in [24]. The rest of this paper is organized as the following. Section 2 briefly describes the server-centric PON-based DCN design. Section 3 shows the completion time and power consumption results, while Section 4 provides the conclusions and future work.

## 2. SERVER-CENTRIC PON-BASED DATA CENTER DESIGN

The server-centric PON-based DCN introduced in [24] and depicted in Figure 1 connects each "cell" with an OLT port in one of its chassis. A 1:N AWGR utilizes a single wavelength to connect the OLT to N "racks" within the cell through one of N groups located in each rack. Servers of all groups within the rack can be interconnected through a passive polymer backplane. Each of the other (N-1) groups utilizes a different wavelength to connect to a group in one of the other (N-1) racks through a splitter. This arrangement allows servers containing non-tunable transceivers to relay traffic through a large number of paths to other servers.

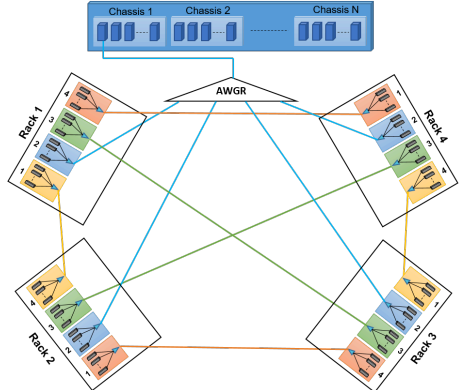

*Figure 1: One cell of the server-centric PON-based DCN design with servers organized in 4 racks and 4 groups per rack [24].*

## 3. OPTIMIZING THE ROUTING FOR MAP-REDUCE SHUFFLING OPERATIONS

### 3.1 Methodology

Mixed Integer Linear Programming (MILP) models are utilized to optimize the routing of intermediate data while balancing the completion time and energy efficiency of MapReduce shuffling. The considered DCNs models are as in [18], in addition to a model for PON option 5 in [24] with 4 racks and 1 server per group. Ten servers are assigned for map and six for reduce as depicted in Figure 2 which results in symmetric 10-to-6 intermediate data flows for the Indy GraySort benchmark [33]. The power consumption values of the selected electronic and optical equipment are summarized in Tables 1 and 2 in [18]. For the PON equipment, the values are taken from [23]. An ON/OFF power profile is assumed.

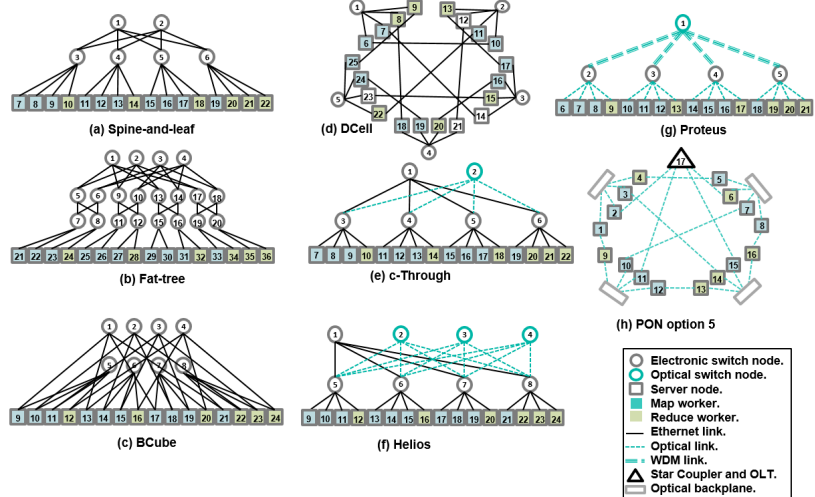

*Figure 2: DCN models considered for MapReduce shuffling optimization.*

### 3.2 Results

Figure 3 summarizes the shuffling completion time results for input data ranging from 1 to 20 GBytes and at maximum server data rate values of 100, 300, 750, and 1000 Mbytes/s. Compared to the results in [18], the server centric PON-based DCN attained an equivalent performance to DCell which achieved the best performance. Figure 4 provides the average networking equipment power consumption results for different maximum server rate values. Compared to Fat-tree, an average reduction by 85% in the power consumption is achieved. While having an equivalent performance to DCell, a power consumption reduction of 67% is gained by the PON-based DCN. Also, power consumption reduction of around 50% is achieved compared to MEMS-based DCNs and Proteus with superior performance.

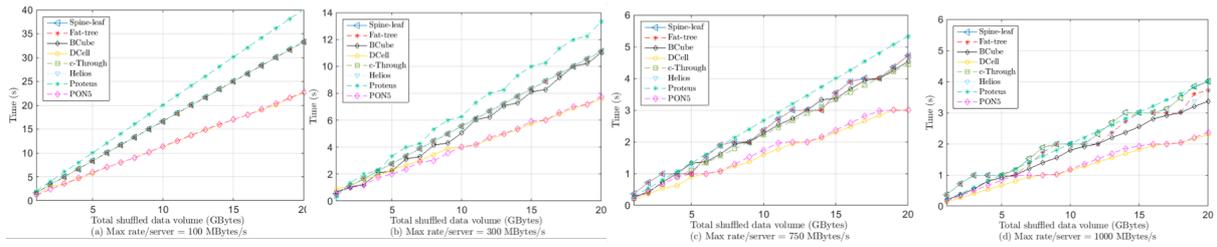

*Figure 3: Shuffling completion time for different DCNs and maximum rate/server.*

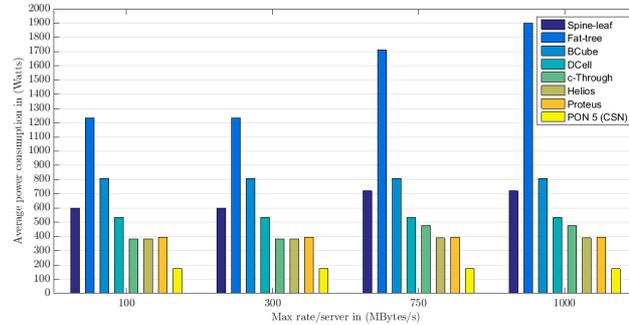

*Figure 4: Average networking equipment power consumption for different maximum rate/server values.*

## 4. CONCLUSIONS AND FUTURE WORK

In this paper, we optimized the completion time and energy efficiency of MapReduce shuffling in a server-centric PON-based DCN and compared the results to other electronic, hybrid and all-optical DCNs. Future work includes considering the switching delays and scalability in DCNs.

## ACKNOWLEDGEMENTS

The authors would like to acknowledge funding from the Engineering and Physical Sciences Research Council (EPSRC), through INTERNET (EP/H040536/1) and STAR (EP/K016873/1) projects. The first author would like to acknowledge EPSRC for funding her PhD programme of study. All data are provided in full in the results section of this paper.

## REFERENCES

[1] K. Dolui and S. K. Datta, "Comparison of edge computing implementations: Fog computing, cloudlet and mobile edge computing," 2017 Global Internet of Things Summit (GIoTS), Geneva, 2017, pp. 1-6.
[2] A. C. Baktir, A. Ozgovde and C. Ersoy, "How Can Edge Computing Benefit From Software-Defined Networking: A Survey, Use Cases, and Future Directions," in *IEEE Communications Surveys & Tutorials*, vol. 19, no. 4, pp. 2359-2391, Fourth quarter 2017.
[3] "Fog Computing and the Internet of Things: Extend the Cloud to Where the Things Are," White Paper, Cisco, 2015.
[4] M. Satyanarayanan, "The Emergence of Edge Computing," in *Computer*, vol. 50, no. 1, pp. 30-39, Jan. 2017.
[5] A. Q. Lawey, T. E. H. El-Gorashi and J. M. H. Elmirghani, "Distributed Energy Efficient Clouds Over Core Networks," in *Journal of Lightwave Technology*, vol. 32, no. 7, pp. 1261-1281, April1, 2014.
[6] L. Nonde, T. E. H. El-Gorashi and J. M. H. Elmirghani, "Energy Efficient Virtual Network Embedding for Cloud Networks," in *Journal of Lightwave Technology*, vol. 33, no. 9, pp. 1828-1849, May1, 1 2015.
[7] A. M. Al-Salim, A. Q. Lawey, T. El-Gorashi and J. M. H. Elmirghani, "Energy Efficient Tapered Data Networks for Big Data processing in IP/WDM networks," *2015 17th International Conference on Transparent Optical Networks (ICTON)*, Budapest, 2015, pp. 1-5.
[8] A. M. Al-Salim, H. M. M. Ali, A. Q. Lawey, T. El-Gorashi and J. M. H. Elmirghani, "Greening big data networks: Volume impact," *2016 18th International Conference on Transparent Optical Networks (ICTON)*, Trento, 2016, pp. 1-6.
[9] A. M. Al-Salim, T. E. El-Gorashi, A. Q. Lawey, and J. M. Elmirghani, "Greening big data networks: velocity impact," IET Optoelectronics, November 2017, DOI: 10.1049/iet-opt.2016.0165.
[10] A. M. Al-Salim, A. Q. Lawey, T. E. H. El-Gorashi, and J. M. H. Elmirghani, "Energy Efficient Big Data Networks: Impact of Volume and Variety," IEEE Transactions on Network and Service Management, vol. PP, no. 99, pp. 1–1, 2017.
[11] H. A. Alharbi, T. E. H. El-Gorashi, A. Q. Lawey and J. M. H. Elmirghani, "Energy efficient virtual machines placement in IP over WDM networks," *2017 19th International Conference on Transparent Optical Networks (ICTON)*, Girona, 2017, pp. 1-4.

[12] A. Hammadi and L. Mhamdi, "A survey on architectures and energy efficiency in data center networks," Computer Communications, vol. 40, pp. 1 – 21, 2014.
[13] László Gyarmati and Tuan Anh Trinh. 2010. How can architecture help to reduce energy consumption in data center networking?. In *Proceedings of the 1st International Conference on Energy-Efficient* Computing and Networking (e-Energy '10). ACM, New York, NY, USA, 183-186.
[14] J. Dean and S. Ghemawat, "MapReduce: Simplified Data Processing on Large Clusters," Commun. ACM, vol. 51, no. 1, pp. 107–113, Jan. 2008.
[15] Y. Shang, D. Li, J. Zhu, and M. Xu, "On the Network Power Effectiveness of Data Center Architectures," Computers, IEEE Transactions on, vol. 64, no. 11, pp. 3237–3248, Nov 2015.
[16] R. Xie and X. Jia, "Data Transfer Scheduling for Maximizing Throughput of Big-Data Computing in Cloud Systems," Cloud Computing, IEEE Transactions on, vol. PP, no. 99, pp. 1–1, 2015.
[17] Z. Kouba, O. Tomanek, and L. Kencl, "Evaluation of Datacenter Network Topology Influence on Hadoop MapReduce Performance," in 2016 5th IEEE International Conference on Cloud Networking (Cloudnet), Oct 2016, pp. 95–100.
[18] S. H. Mohamed, T. E. H. El-Gorashi and J. M. H. Elmirghani, "On the energy efficiency of MapReduce shuffling operations in data centers," *2017 19th International Conference on Transparent Optical Networks (ICTON)*, Girona, 2017, pp. 1-5.
[19] Y. Gong, Y. Lu, X. Hong, S. He and J. Chen, "Passive optical interconnects at top of the rack for data center networks," *2014 International Conference on Optical Network Design and Modeling*, Stockholm, 2014, pp. 78-83.
[20] Kachris and I. Tomkos, "Power consumption evaluation of hybrid WDM PON networks for data centers," in 2011 16th European Conference on Networks and Optical Communications, July 2011, pp. 118–121.
[21] Y. Cheng, M. Fiorani, R. Lin, L. Wosinska, and J. Chen, "POTORI: a passive optical top-of-rack interconnect architecture for data centers," IEEE/OSA Journal of Optical Communications and Networking, vol. 9, no. 5, pp. 401–411, May 2017.
[22] J. Elmirghani, T. EL-GORASHI, and A. HAMMADI, "Passive optical-based data center networks," 2016, wO Patent App. PCT/GB2015/053,604.
[23] A. Hammadi, T. E. H. El-Gorashi, and J. M. H. Elmirghani, "high performance AWGR PONs in data centre networks," in 2015 17th International Conference on Transparent Optical Networks (ICTON), Budapest, 2015, pp. 1-5.
[24] A. Hammadi, T. E. H. El-Gorashi, M. O. I. Musa and J. M. H. Elmirghani, "Server-centric PON data center architecture," *2016 18th International Conference on Transparent Optical Networks (ICTON)*, Trento, 2016, pp. 1-4.
[25] R. Alani, A. Hammadi, T. E. H. El-Gorashi and J. M. H. Elmirghani, "PON data centre design with AWGR and server based routing," *2017 19th International Conference on Transparent Optical Networks (ICTON)*, Girona, 2017, pp. 1-4.
[26] A. Hammadi, M. Musa, T. E. H. El-Gorashi and J. H. Elmirghani, "Resource provisioning for cloud PON AWGR-based data center architecture," *2016 21st European Conference on Networks and Optical Communications (NOC)*, Lisbon, 2016, pp. 178-182.
[27] A. Hammadi, T. E. H. El-Gorashi and J. M. H. Elmirghani, "Energy-efficient software-defined AWGR-based PON data center network," *2016 18th International Conference on Transparent Optical Networks (ICTON)*, Trento, 2016, pp. 1-5.
[28] J. Beals IV, N. Bamiedakis, A. Wonfor, R. Penty, I. White, J. DeGroot Jr, et al.: A terabit capacity passive polymer optical backplane based on a novel meshed waveguide architecture, Applied Physics A, vol. 95, pp. 983-988, 2009.
[29] A. S. Thyagaturu, A. Mercian, M. P. McGarry, M. Reisslein and W. Kellerer, "Software Defined Optical Networks (SDONs): A Comprehensive Survey," in *IEEE Communications Surveys & Tutorials*, vol. 18, no. 4, pp. 2738-2786, Fourth quarter 2016.
[30] Y. Luo, F. Effenberger and M. Sui, "Cloud computing provisioning over Passive Optical Networks," *2012 1st IEEE International Conference on Communications in China (ICCC)*, Beijing, 2012, pp. 255-259.
[31] M. Taheri and N. Ansari, "A feasible solution to provide cloud computing over optical networks," in IEEE Network, vol. 27, no. 6, pp. 31-35, November-December 2013.
[32] S. H. S. Newaz, W. Susanty binti Haji Suhaili, G. M. Lee, M. R. Uddin, A. F. Y. Mohammed and J. K. Choi, "Towards realizing the importance of placing fog computing facilities at the central office of a PON," *2017 19th International Conference on Advanced Communication Technology (ICACT)*, Bongpyeong, 2017, pp. 152-157.
[33] A. Rasmussen, G. Porter, M. Conley, H. V. Madhyastha, R. N. Mysore, A. Pucher, and A. Vahdat, "TritonSort: A Balanced and Energy Efficient Large-Scale Sorting System," ACM Trans. Comput. Syst., vol. 31, no. 1, pp. 3:1–3:28, Feb. 2013.